# Dynamics of Molecular Clouds

by

George B. Field, Eric G. Blackman, and Eric R. Keto

## Abstract

We further develop the model of molecular cloud fragmentation introduced in Field, Blackman and Keto (2007; FBK). Here we show that external pressure $P_e$ acting on fragments establishes a scale-dependent critical mass, $M_c \propto 1/P_e$ when incorporated into the time dependent virial theorem. Fragments with $M < M_c$ are confined largely by pressure, while those with $M \geq M_c$ collapse under self gravitation. Both types of fragments are commonly observed. Without specifying the source of $P_e$, and without assuming any other scaling relations, we first predict the observed power law index $p_1$ in the relation $\sigma = V_0 L^{p_1}$ that connects the rms velocity $\sigma$ of supersonic motions to the size of fragments $L$. We then investigate the possibility that $P_e$ is due to the kinetic energy of H atoms released by photodissociation of H₂ in the fragment. This can account approximately for observed values of $P_e$ and two additional observations: the value of the coefficient $V_0$ above, and the observation of outflowing atomic H around MCs. A further prediction is HI at fragment edges with column densities of order $10^{20}$ cm$^{-2}$ and velocities of a few km/s that should be detectable with high resolution 21 cm observations. Finally, we predict the magnitude of the coefficient of dissipation in the observed supersonic flows.

*1. Introduction*

Complementing recent simulations of collapsing molecular clouds (Va'zquez-Semadini et al 2007, 2009), Field, Blackman, and Keto (2007;FBK) presented an analytic model of the fragmentation of molecular clouds in which the observed macroscopic velocities are driven solely by gravitational energy released in the fragmentation process. FBK derived scaling exponents for cloud properties like mass and number in terms of the exponent $p_1$ for the dependence of the velocity dispersion $\sigma$ on the size $L$ of the fragment. They adopted the observationally inferred value of $p_1 = \frac{1}{2}$ without a theoretical derivation (Table 2 of FBK). Here we show that an additional physical hypothesis not only explains why $p_1 = \frac{1}{2}$, but predicts a scaling coefficient which is in approximate agreement with observation.

Our discussion in § 2 is based on the time – dependent virial theorem for spherical fragments immersed in a constant external pressure $P_e$. In § 3 we derive a scale - dependent critical mass $M_c$, above which structures collapse. In § 4 we discuss the physical importance of this mass for the molecular cloud fragmentation cascade and show that it helps to explain the data of Keto and Myers (1986) and Bertoldi and McKee (1992). In § 5 we show that if the masses of all fragments in the cascade are about equal to $M_c$, then $p_1 = \frac{1}{2}$ and we can calculate the coefficient in the scaling relation. In section § 6 we discuss the possibility that $P_e$ is due to energetic H atoms resulting from the photodissociation of molecules by far UV photons. A model of this process is summarized in Appendix 1 and further discussed in Field et al (2010). In Appendix 2 we discuss other possible contributions to the external pressure. In Appendix 3 we consider the dissipation of energy in the supersonic flow of the cascade. We show that the dissipation is governed by the ratio of a cloud's free fall time its energy dissipation time, and that this ratio, $\gamma$, is equal to $p_1 = \frac{1}{2}$. Our conclusions are discussed in § 7.

## 2. Time Dependent Virial Theorem

For spherical fragments of mass $M$ and radius $R$ bounded by a uniform external pressure $P_e$ the virial theorem is

$$\frac{\ddot{I}}{M} = 3\sigma^2 - \frac{\Gamma GM}{R} - \frac{4\pi R^3}{M} P_e = R\left(\frac{3\sigma^2}{R} - \frac{\Gamma GM}{R^2} - \frac{4\pi R^2}{M} P_e\right). \tag{1}$$

Note that in the second expression $M$ enters only in the combination $M/R^2 = \pi\Sigma = \pi\mu m_H N(H_2)$, where $\Sigma$ is the mass per unit area, and $N$ is the column density. Here $\mu = 2.33$ is the mean molecular weight per molecule. We note that $N$ is directly observable. The solutions to (1) for $\sigma^2/L$ in terms of $\Sigma$ are shown in Figure 1.

For simplicity, we assume that fragments are isothermal spheres, for which the form factor $\Gamma$ in (1) is tabulated by Elmegreen (1989). For reasons to be explained later, we take $\Gamma = 0.73$. In what follows we shall be interested only in the sign, not the magnitude of $\ddot{I}$. The reason is that observations cannot determine $\ddot{I}$, only its average value $T^{-1}\int \ddot{I}dt = \dot{I}/T$ over an appropriate time interval $T$. If the sign of $\dot{I}$ is correlated with that of $\ddot{I}$, then the sign of each indicates whether the fragment is collapsing (-), in equilibrium (0), or expanding (+). For given values of $\sigma$, $M$ and $P_e$, $\ddot{I}$ is a function of $R$. The third term dominates the compression at small $R$ and the fourth term dominates at large $R$, suggesting that there is a critical value of $R$, $R_c$, at which compression is maximal. If the absolute value of the compressive terms at $R_c$ is greater than $3\sigma^2$, equilibrium is not possible. We discuss this situation further in § 2.

If we set $P_e = 0$ in equation (1) and look for equilibrium solutions, we conclude that $M = 3\sigma^2 R/\Gamma G$, which is sometimes interpreted as the "virial mass", denoted $M_V$. Elmegreen (1989) shows that this is legitimate if the configuration naturally has internal pressure which goes to zero at the boundary, as for polytropes of certain indices. However, observers find that finite isothermal spheres provide good fits to the observations of MCs. Since only isothermal spheres with infinite radii and masses can have $P_e = 0$ we infer that $P_e \neq 0$ for real clouds.

In some observations, the actual value of $P_e$ can be found, as in a landmark paper by Keto and Myers (1986), who found that $P_e/k = 3E3$ to $3E4$ cm$^{-3}$ K for structures at high galactic latitude. Bertoldi and McKee (1992) showed that observations of MCs in Ophiuchus can be understood if $P_e/k$ is within a factor of two of $10^5$. We wish to determine if a similar explanation applies more generally, and therefore adopt isothermal spheres with finite $P_e$ as a model for fragments. Chièze (1987) explored the consequences of such a model, as had Whitworth and Summers (1985).

3. *Critical Mass*

The concept of critical mass is important in understanding fragmentation. We start with a key insight by Chièze, that the existence of a finite external pressure enables one to write the radius $R$ in a nondimensional form:

$$X = G^{-1/4} P_e^{1/4} M^{-1/2} R, \qquad (2)$$

in terms of which (1) can be written

$$\frac{\ddot{I}}{M} = 3\left[\sigma^2 - f(X) G^{3/4} P_e^{1/4} M^{1/2}\right], \qquad (3)$$

where

$$f(X) = \frac{1}{3}\left(\frac{\Gamma}{X} + 4\pi X^3\right). \qquad (4)$$

$f(X)$ has a minimum, so as mentioned in Section 2, $\ddot{I}$ has a maximum at a critical value of $X$, for which Chieze gets $X_c = (8\pi)^{-1/4} = 0.447$ with a derivation that does not use the equation of virial equilibrium. For reasons that will become clear below, we will assume approximate virial equilibrium for collapsing fragments, and further, that in Nature the variable $X$ in (4) takes a critical value that is an extremum of $f(X)$. According to (4) this occurs at the critical value of $X$ given by

$$X_c = \left(\Gamma/12\pi\right)^{1/4} = 0.37, \qquad (5)$$

where we use $\Gamma = 0.73$. To maintain consistency, we shall use this value rather than that of Chièze stated above.

Using (4) and (5) we find that

$$f(X_c) = 0.87. \qquad (6)$$

We conclude that for all values of $M$ such that

$$0.87 G^{3/4} P_e^{1/4} M^{1/2} > \sigma^2, \qquad (7)$$

$$\ddot{I} < 0, \qquad (8)$$

so collapse must take place. In other words, collapse is inevitable if $M \geq M_c$, where $M_c$ is a critical mass given by

$$M_c = 1.3 \frac{\sigma^4}{G^{3/2} P_e^{1/2}}. \qquad (9)$$

We can understand this qualitatively; both $\sigma$ and $1/P_e$ tend toward expansion, but if $M$ is large enough, self gravitation dominates. Note that Mouschovias and Spitzer (1976) derive the same formula differently, obtaining 1.2 rather than 1.3.

### 4. *How Fragmentation Works*

In the model of FBK, fragments with $M \geq M_c$ collapse and produce fragments of smaller $M$. This process continues until fragments with $M < M_c$ are produced, at which point $\ddot{I} \geq 0$ and contraction is no longer inevitable. In some cases, fragments emerge with $\ddot{I} \cong 0$, signifying equilibrium. Bonnor (1956) – Ebert (1957) spheres are isothermal spheres in such equilibrium. As discussed by FBK, equilibrium spheres fall on a curve with two branches in the $P_e - R^3$ plane. Spheres on the large $R$ (pressure or P) branch are confined mostly by pressure and are stable. Those on the small $R$ (gravitational or G) branch are confined mostly by gravitation, and are unstable to collapse.

   Keto and Myers (1986) studied high – latitude molecular clouds (HLCs), and found that their masses are too low to account for a significant contribution to confinement by self gravitation. They compared their data with those of larger column density, as indicated by Figure 1. Using the virial theorem and including pressure (see (1)), they showed that the HLCs can be confined by an external pressure if it is in the range $P_e/k = 3E4$ to $3E4$ cm$^{-3}$ K. This contrasts with the more massive clouds near the Galactic plane, which are often gravitationally bound, and may be associated with the fact that confining pressure far from the plane may be lower than that near the plane (Field et al 2010).

   Bertoldi and McKee (1992) found that in Ophiuchus, a well - studied molecular cloud near the galactic plane, the data indicate that there are two classes of what they referred to as clumps (which in our fragmentation cascade are simply fragments below ~10pc in size). They found that those with masses smaller than $40 M_\odot$ are not gravitationally bound, while those of larger masses are. They showed that the small – mass fragments can all be confined by external pressure if $P_e/k \sim E5$ cm$^{-3}$ K within a factor of two.

The larger – mass objects all have $M \cong M_c$ appropriate for their observed values of $\sigma$, and are therefore gravitationally confined.

These facts can be interpreted within our framework. To do so, imagine a gravitational fragmentation cascade taking place, with $M \geq M_c$, as envisioned by FBC. From time to time, the fragmentation process may result in a fragment with $M < M_c$. Such a fragment can come into equilibrium with the prevailing pressure in the region. If so, it would be on the P branch, on which the equilibrium is stable. We identify the Ophiuchus and HLC fragments of low mass with such objects. It is consistent with this interpretation that no fragments with masses < 40 $M_\odot$ are observed in Ophiuchus to have $M > M_c$.

On the other hand, the more massive fragments are observed to obey $M \cong M_c$, so they must be collapsing as part of the gravitational cascade. Why are there few fragments with $M >> M_c$? To address this question, let the number of fragments produced at each step be $N$, so that $M_2 = M_1 / N$, and $\log(M_2 / M_1) = -\log N$. $N$ should not be very large, because the first length scale within a collapsing fragment that becomes gravitationally unstable is the largest one that is significantly smaller than the collapsing fragment, so we estimate that $2 < N < 7$, say $5$, in which case log ($M_2 / M_1$) = $-0.7$. This may be compared to the total range of observed masses, $\Delta \log M = -6$. Thus the gravitational cascade, like that in incompressible turbulence, is local in the sense that it is rare for a clump to directly fragment into many much smaller ones. Thus, we may assume that the daughters of a fragment with $M \simeq M_c$ at their initial scale also have $M \simeq M_c$ appropriate for the daughter scale.

To pursue these ideas we introduce a new hypothesis: fragments with masses $M < M_c$ are primarily pressure confined and are stable against collapse, while those with masses $M \simeq M_c$ fragment until $M \simeq M_c$ again. We therefore predict that all fragments on the G branch have masses $M$ that are about equal to the critical value, $M_c$, and are therefore marginally unstable. This hypothesis is consistent with the discussion in FBK. It has observable consequences, as we show below.

### 5. Using $M \cong M_c$ to *predict the velocity–size scaling relation*

From (2) and the assumption that $X = X_c$ we see that $M \propto R^2$, so that from (9) $\sigma \propto L^{\frac{1}{2}}$, or $p_1 = \frac{1}{2}$, a relationship that FBK previously only empirically inferred from observations without a theoretical derivation. Observations of MCs by Heyer and Brunt (2004) yield $p_1 = 0.6 \pm 0.07$, not far from $\frac{1}{2}$. The present derivation of $p_1 = \frac{1}{2}$ has not required any other observational scaling relations (i.e. we are not assuming one of Larson's laws to obtain the other). Instead, it results from three physical principles : (i) that $\ddot{I} < 0$ is the condition for collapse (ii) that many observed fragments require a bounding external pressure for equilibrium and (iii) that there is a critical mass above which the maximum value of $\ddot{I}$ is negative. The equation for $M_c$ (or $X = X_c$) and the physical arguments that structures evolve toward $M_c$ eliminate the need for assuming any other scaling relations.

We can go further and constrain the coefficient in the scaling velocity-size relationship from our present work. For this purpose we write

$$\sigma = V_0 L_{pc}^{1/2} \text{ km s}^{-1}, \tag{10}$$

following Heyer and Brunt (2004). Equation (3) with $X = X_c$ can be written in the form

$$\frac{\sigma^2}{L} = \left(\frac{2\pi}{3} X_c^2 + \frac{\Gamma}{6 X_c^2}\right)(GP_e)^{1/2} = 1.16(GP_e)^{1/2}, \tag{11}$$

where we have used $\Gamma = 0.73$ and $X_c^2 = 0.14$. This formula confirms that $p_1 = \tfrac{1}{2}$, and also allows us to calculate the scaling coefficient $V_0$:

$$V_0^2 = 3.5E-12\left(\frac{P_e}{k}\right)^{1/2} \text{ cm s}^{-2} = 1.08E-3\left(\frac{P_e}{k}\right)^{1/2} \text{ km}^2 \text{ s}^{-2} \text{ pc}^{-1}. \tag{12}$$

We come back to (12) in the next section.

### 6. External Pressure

Although the existence of external pressure is accepted, there has not yet been agreement as to its nature. Presumably it is due to gas in some way, but if so, is the hydrogen in one of the familiar phases – HI, HII or $H_2$? Is the pressure due to macroscopic or to therrnal motions? Here we investigate the possibility that it is due to atomic hydrogen at the edge of the MC, produced and heated by the dissociation of $H_2$ in the MC by far

UV photons. A detailed calculation of this effect is given in Appendix 1. In Appendix 2 we describe other possible contributors to $P_e$. Here we summarize the key points.

There is an extensive literature on photodissociation regions (PDRs), which form at the edges of MCs that are exposed to photons in the range 91.2 to 111.0 nm (Hollenbach and Tielens 1999). These so - called Far UV photons can dissociate $H_2$ molecules, but cannot ionize HI atoms, so they can reach MCs through intervening HI. PDRs emit a characteristic spectrum, which has been studied by many groups, allowing one to infer temperatures, compositions, and densities of many PDRs. Naturally, they tend to be found near OB stars, the sources of most FUV photons. We give a model for the ensuing outflow of H atoms in Appendix 1. We find that for realistic estimates of the FUV intensity, the heated HI results in a significant pressure at the edge of an MC which we refer to as photodissociation pressure. Each FUV photon deposits 0.13 eV in each HI atom. If we assume that this energy is shared with the surrounding gas, the resulting pressure serves to help confine the MC, and also accelerates the atomic H to high enough speed that it leaves the cloud.

Many PDRs have been analyzed, and the appropriate values of $\chi$, a dimensionless measure of the far UV intensity described in Appendix 1, have been derived. The results range from a minimum of 2 to a maximum of 200. Cubick et al (2008) found that the far infrared emission of the Galaxy can be explained if a mean value for the Galaxy of $\chi = 60$ is adopted.

If $P_e$ results from photodissociation pressure, then (12) can be written

$$V_0 = 0.033 \left(\frac{P_e}{k}\right)^{1/4} = 0.21 \chi^{1/4} \text{ km s}^{-1} \text{pc}^{-1/2}, \qquad (13)$$

According to Heyer and Brunt (2004), $V_0 = 0.8 - 1.2$ km s$^{-1}$pc$^{-1/2}$. With Cubick's average value $\chi = 60$, (13) gives $V_0 = 0.6$. It also gives $P_e/k = 1E5$ cm$^{-3}$ K, in the range found by Bertoldi and McKee for the pressure - bound fragments in Ophiuchus, and is more than adequate to explain the HLC's observed by Keto and Myers. We refer the reader to an application of external pressure to other regions in Field et al (2010).

## 7. Conclusions

We have further developed the FBK model of MC fragmentation by investigating the dynamics in greater detail. We posit the existence of an external pressure $P_e$ which contributes along with gravitation to the confinement of MCs.

In the context of confinement by both gravitation and pressure we identify a critical mass, $M_c$, that is proportional to $1/P_e^{1/2}$. We show that fragmentation, along with the hypothesis that fragments in the cascade are marginally unstable and therefore satisfy $M \simeq M_c$, can explain the HLC data of Keto and Myers (1986) and those of Bertoldi and McKee (1992) on fragments in Ophiuchus. The same hypothesis also predicts both the observed scaling exponent of $\sigma$ with $L$, the size of the fragment, and also predicts the value of the scaling coefficient.

We also studied the possibility that the external pressure is due to the kinetic energy of H atoms released by photodissociation of $H_2$ molecules by FUV radiation. A simple model of this process yields external pressures that would contribute significantly, along with self gravitation, to confinement. Our basic prediction of Larson's laws from the use of an external pressure and our derived critical mass on each scale does not however depend on the mechanism of external confinement.

*Appendix 1: A Model of Dissociation Pressure*

Each dissociation event results in a kinetic energy $E$ being deposited in each H atom. According to Stephens and Dalgarno (1973) $E = 2.1E-13$ erg. To proceed, we need to know the intensity of FUV photons $\Phi$, which is proportional to a quantity $\Im$, which Habing (1968) calculated to be $\Im = 1.2 \times 10^7$ ph cm$^{-2}$ s$^{-1}$ in the solar neighborhood. He pointed out that substantially larger values are expected in the vicinity of OB associations because of the large FUV fluxes from such stars. We accommodate that fact by introducing a dimensionless parameter $\chi$ defined by

$$\Im = 1.2E7 \chi \text{ ph cm}^{-2}\text{s}^{-1}. \tag{14}$$

The flux of photons incident upon the surface of a fragment, $\Phi$, is related to $\Im$ by

$$\Phi = a_1 a_2 a_3 a_4 \Im, \tag{15}$$

where

$$a_1 = 0.5 \tag{16}$$

accounts for the fact that because of absorption in the fragment itself, virtually no photons arrive at the surface from the direction of the fragment in question,

$$a_2 = 0.5 \tag{17}$$

accounts for the fact that the net flux toward the fragment is half of the intensity at the surface,

$$a_3 = 0.5 \tag{18}$$

accounts for the fact that only half of the FUV photons are absorbed by $H_2$ rather than by dust (Draine and Bertoldi 1996) and

$$a_4 = 0.13 \tag{19}$$

is the fraction of FUV photons absorbed by $H_2$ which result in dissociation (Draine and Bertoldi 1996). Therefore

$$\Phi = 2E5\chi \text{ ph cm}^{-2}\text{s}^{-1}. \tag{20}$$

Various observers, including Stutzki et al (1988), Schneider et al (1998), Kulesa et al (2005), Pinada et al (2008) and Sun et al (2008), have calculated the values of $\chi$ needed to explain the data in various PDRs. They range from 2 to 200, with an average of 110. Cubick et al (2008) aimed to explain the far infrared radiation from the Galaxy observed by COBE. They found that most of such radiation originates in PDRs, and that the best fit to their data is $\chi = 60$.

The effect of the hot atoms is to increase the pressure at the surface of the fragment. We assume that the FUV radiation is absorbed in a thin region at the surface of the fragment. The pressure there is in equilibrium with the external pressure $P_e$ in the fragment. The pressure far outside the fragment is assumed to be negligible, so the pressure gradient drives an outward flow of atoms, which is one of the predictions of the model. According to Bernoulli's law,

$$P_e = \rho v^2, \tag{21}$$

where $\rho$ is the density and $v$ is the speed at the exit of the flow of atoms from the fragment. We denote the constant value of the mass flux by

$$\rho v = 2m\Phi, \tag{22}$$

so from (21)

$$P_e = 2m\Phi v. \tag{23}$$

We denote the fraction of the energy $E$ that ends up in the kinetic energy of the flow by $\xi$, a parameter that can be found only by a study of radiation losses which we do not attempt here. Then the flux of energy available for acceleration of the flow is $\Phi \varepsilon E$. The resulting flux of kinetic energy is $\tfrac{1}{2}\Phi m v^2$. Equating the two fluxes results in

$$v = \left( \frac{2\xi E}{m} \right)^{1/2}. \tag{24}$$

Putting (24) into (23) gives

$$P_e = 2\Phi \left( 2\xi m E \right)^{1/2}, \tag{25}$$

so, using (20), we find that

$$\frac{P_e}{k} = 2400 \xi^{1/2} \chi \text{ cm}^{-3} \text{ K}. \tag{26}$$

If we take $\xi = \tfrac{1}{2}$ as a reasonable guess, we get $1700\chi$, the value used to get (13) and the following results cited in the text.

*Appendix 2: Other Contributors to the External Pressure*

As we have seen, the existence of a constant external pressure offers a way to understand the velocity scaling and the evolution of fragments in the FBK model. Although we have proposed that the required pressure is due to dissociations, this hypothesis has not yet been confirmed. Here we consider other possible contributors to the pressure.

From an observational point of view, pressures are different in the diffuse ISM (composed of HI), ionized regions around hot stars (composed of HII) and MCs (composed of $H_2$). The temperatures and pressures in the diffuse ISM are determined by 21 – cm observations, as well as by observations of ultraviolet absorption lines in spectra of early – type stars. It is found that typical thermal pressures are 3000 $cm^{-3}$ K (Jenkins and Tripp 2007), less than is needed in many MCs. Although macroscopic motions of HI account for a turbulent pressure of about $2E4$ $cm^{-3}$ K (Elmegreen 1989), this value is still too small to confine MCs near the galactic plane. HII regions are not pervasive enough to play a role in most MCs, (but see below).

Keto and Myers (1986) invoked values consistent with Elmegreen's later (1989) discussion to account for the confinement of their observed high – latitude MCs (HLCs), which are isolated structures far from the galactic plane. The masses of HLCs are so low that self gravitation is clearly not enough to confine them. At least in the case of the Ophiuchus clumps studied by Bertoldi and McKee (1992), Elmegreen's relatively low values of the pressure are not enough to explain them.

Another possibility is that fragments are confined by turbulence in some medium within the parent MC. Against this hypothesis, Ballesteros - Paredes et al (2006) showed that turbulent motions at the surface of a gaseous structure tend not to confine it, but rather to disrupt it. However, even if we assume that such pressure is important in confining fragments, its value $\rho_e v_t^2$, where $\rho_e$ is the density of the confining gas and $v_t$ is its rms turbulent velocity, is subject to constraints. In the FBK model the observed motions of fragments $\sigma$ are driven by self gravitation, and may not be turbulent in the conventional sense. However, they give rise to a stress at the surface of the fragment equal to $\rho_s \sigma^2$, where if $M = M_c$, $\rho_s = 0.4\bar{\rho}$ is the density at the surface of a fragment whose mean density is $\bar{\rho}$ (Elmegreen 1989). Then the pressure balance condition at the surface of a fragment is

$$\rho_s \sigma^2 = \rho_e v_t^2 \quad . \tag{27}$$

(27) is also is appropriate for fragments on the P branch, with $\rho_s = \bar{\rho}$. We assume that $\rho_s > \rho_e$ if the fragment is confined by a low – density medium. It is observed that for fragments on the G branch, $\sigma > c_s = 0.2$ km s$^{-1}$, the sound speed, while on the P branch, $\sigma \geq c_s$. Thus all fragments obey $\sigma \geq c_s$. It follows from (27) that

$$v_t > c_s \tag{28}$$

so that the putative confining turbulence must be supersonic with respect to molecular gas. In the FBK model, the observed supersonic motions within fragments are driven by gravitational instability. If the confining gas is molecular and of the same density as the cloud, it would be indistinguishable from the material in the fragments themselves, and the idea of a separate confining medium would not be applicable. If the confining medium could be simply characterized by a lower average density and higher associated velocity dispersion consistent with Larson's laws, then the turbulent pressure would be sufficient in magnitude to support smaller scale clumps. However, the very concept of turbulent pressure support of a small - scale structure by larger turbulent eddies is implausible.  Confinement by a supersonic turbulent molecular gas might be sustained only if there were an independent source of smaller scale turbulent energy, due for example to feedback from stars.

Perhaps the confining gas is HI, as in the dissociation model of Appendix 1, but whose pressure is turbulent rather than thermal. As indicated above, the turbulent motions would have to exceed 0.2 km s$^{-1}$ Such macroscopic motions of HI are indeed observed in the diffuse ISM but not yet in MCs. They are thought to be driven by supernova explosions as in the model of the ISM of McKee and Ostriker (1977). To explain confining pressures, SN explosions would also have to affect the motions of molecular gas, which would not be consistent with our model. Because our purpose here is to explore the consequences of our model, we put aside this hypothesis.

Another possibility is that the external pressure is thermal in nature, with $P_e = \rho c_s^2$, where the sound speed $c_s$ may have a variety of different values to be discussed below. The speed of sound depends upon the temperature of the external gas, which in turn depends upon whether the gas is molecular, atomic, photoionized, or shock - heated by stellar winds or explosions. The respective characteristic temperatures are of the order of 10 K, 100 K, $10^4$ K and $10^6$ K, respectively. Which if any choice may be applicable to the confining gas if $P_e$ is thermal in nature? If the gas is molecular, its temperature would be comparable to that within the fragments, 10 K, and the above constraint shows that this case is unrealistic.

If the gas is atomic, its temperature might be higher, and therefore its density lower, thus avoiding the previous constraint. Indeed this is the case in the model of Appendix 1. To test it, we propose that 21 – cm observations of individual fragments be undertaken. These observations should have high angular resolution to distinguish the proposed surface layer of HI and its outflow from background HI signals originating in the general diffuse ISM. We predict warm HI with speeds of several km s$^{-1}$ and column densities of a few times $10^{20}$ cm$^{-2}$.

Photoionized gas at $10^4$ K, would require a density of 10 cm$^{-3}$. Such gas would be easily detected in observations of radio recombination lines, but it is not , so this possibility is ruled out. Another possibility is shock - heated gas at $10^6$ K. Gödel et al (2007) observed x rays coming from the direction of the Orion MC, and showed that they originate in a diffuse gas with T = $10^6$ K and a pressure $P/k = 1E5$ cm$^{-3}$K, about what is needed. They suggest that the gas is being shock heated by strong winds from stars in the Orion cluster. If this were a general phenomenon, pressure - confined fragments would be found only close to young massive stars, because only such stars have sufficiently powerful winds to heat gas to $10^6$

K. Lacking information on this point, we do not invoke Gödel's phenomenon to confine fragments throughout the Galaxy at this time.

*Appendix 3: Energy Dissipation and the Velocity Scaling Exponent*

Here we show that the value of $p_1$ derived in §5, $\frac{1}{2}$, is required if the dissipation is by shock waves, which must occur in supersonic flows like those observed in MCs. Kolmogorov's (1941) well – known velocity exponent, $p_1 = \frac{1}{3}$, applies only to the inertial range of subsonic turbulence. It depends completely on his assumption that the kinetic energy which cascades to smaller scales by nonlinear interactions throughout the inertial range is conserved because dissipation occurs only on scales smaller than the inertial range by definition.

Kolmogorov's reasoning does not apply to MCs because the internal motions are supersonic, unlike those treated by Kolmogorov. Moreover, in the FBK model, gravitational energy drives the motions, an effect not considered by Kolmogorov. Numerous simulations, such as those reviewed by Elmegreen and Scalo (2004) , show that supersonic flows are dominated by shock waves on all scales. In shock waves there are thin layers where viscous dissipation takes place. Thus motions on all scales dissipate directly, rather than only those at the end of an energy cascade as in subsonic turbulence.

Our discussion of dissipation in gravitationally driven supersonic flow is based on a consideration of the disposition of total energy per unit mass $E$ and kinetic energy per unit mass $K$ as matter cascades to smaller scales in the FBK model. Mass is conserved in the cascade, so it is sufficient to consider only the energies per unit mass. Since $E = K + W$, where $W$ is the gravitational potential energy per unit mass, we immediately recognize a difference from Kolmogoroff's model, in which $W = 0$. We are able to eliminate the variable $W$ because of our key assumption that at each stage in the cascade, the masses of fragments equal the critical mass, in which $W$ is related to $K$. We exploit this fact in what follows.

From (3) and (9)

$$\frac{M}{M_c} = \frac{0.8}{f^2(X)}, \tag{29}$$

which must exceed unity for fragmentation to occur, implying that $X$ must be such that $f(X) \leq 0.9$. This is a strong requirement because this happens only at a single value, $X = X_c$. According to (5), in virial equilibrium this occurs when $X = X_c = \left(\frac{\Gamma}{12\pi}\right)^{1/4} = 0.37$. We adopt this value in what follows. It is convenient to define a new variable,

$$Y = 4\pi X^4 / \Gamma, \tag{30}$$

in terms of which the critical value of $Y$ is

$$Y_c = \frac{1}{3}. \tag{31}$$

The virial theorem (3) for a near – equilibrium structure can now be written in the form

$$2K + W = K + E = \frac{3P_e V}{M} = \frac{2Y}{1+Y} K, \tag{32}$$

where we have used (2), (3) and (30) in deriving the last equality. Hence

$$E = \frac{Y-1}{Y+1} K,  \tag{33}$$

so that $E = 0$ for $Y = 1$, $-K$ for $Y \to 0$, and $+K$ for $Y \to \infty$. For $Y = Y_c = 1/3$,

$$E = -\tfrac{1}{2} K  \tag{34}$$

and

$$\frac{3 P_e V}{M} = \tfrac{1}{2} K .  \tag{35}$$

Since dissipation acts only on the kinetic energy, we assume that $dK = -K dt / t_D$, where $t_D$ is the mean time for shock dissipation, to be discussed further below. According to (34), for a fixed value of $Y = Y_c$ this implies also that $dE = -E dt / t_D$. Now consider the effect of dissipation on the conservation of energy. This is stated in terms of $E_L$, the energy per mass per unit length. Since near virial equilibrium applies to every scale, the foregoing equations involving $E$ and $K$ also apply to $E_L$ and $K_L$. Therefore the equation of energy conservation is

$$\partial_t E_L + \partial_L (E_L \dot{L}) = -E_L / t_D.  \tag{36}$$

We define the energy flux per unit mass $\varepsilon$ by

$$\varepsilon = E_L \dot{L},  \tag{37}$$

so that in a steady-state fragmentation cascade that occurs for $Y = \tfrac{1}{3}$, (36) can be written in the form

$$L\frac{d\varepsilon}{dL} = -\frac{\varepsilon}{t_D}\frac{L}{\dot{L}}, \tag{38}$$

or

$$\frac{d\ln\varepsilon}{d\ln L} = \frac{t_G}{t_D} \equiv \gamma. \tag{39}$$

Here we have used the fact that as a result of fragmentation,

$$\dot{L} = -L/t_G, \tag{40}$$

where

$$t_G = \left[4\pi G \bar{\rho}(L)\right]^{-1/2} \tag{41}$$

is the gravitational instability time scale at the scale $L$ in terms of the mean density at that scale. We show in (52) below that

$$t_G \propto L/\sigma(L), \tag{42}$$

and since the time scale for shock dissipation is also proportional to the crossing time $L/\sigma(L)$, the dissipation parameter $\gamma$ is independent of $L$. Therefore, (39) implies that

$$\varepsilon \propto L^\gamma. \tag{43}$$

Moreover, it follows from (40) that

$$\dot{L} \propto \sigma. \tag{44}$$

But from (37)

$$\varepsilon = \dot{L}E_L \propto \sigma K_L \propto \sigma^3/L \propto L^{3p_1-1}, \tag{45}$$

so from (43) we conclude that

$$p_1 = \frac{1+\gamma}{3}. \tag{46}$$

It follows from $p_1 = \tfrac{1}{2}$ (see §5) that

$$\gamma = 3p_1 - 1 = \tfrac{1}{2}. \tag{47}$$

With $t_G = [4\pi G \bar{\rho}(L)]^{\tfrac{1}{2}}$ we can write

$$\left(\frac{\sigma t_G}{R}\right)^2 = \frac{\sigma^2 R}{3GM} = \frac{2\Gamma K}{9(-V)} = \frac{2\Gamma K}{9(K-E)} = \frac{\Gamma(Y_c + 1)}{9}, \tag{48}$$

so

$$t_G = \frac{[\Gamma(1+y_c)]^{\tfrac{1}{2}}}{6}\frac{L}{\sigma} = 0.29\frac{L}{\sigma_{3D}} = \gamma t_D, \tag{49}$$

where the last equality follows from (39) and $\sigma_{3D} = \sqrt{3}\sigma$. It follows that if we can determine $\gamma$ independently, we can compute the necessary value of $t_D$ from the value of the free – fall time, which is a simple function of the density as a function of scale, as indicated in (41).

(46) has interesting implications. The first is that if there is no dissipation in the cascade, as in the Kolmogoroff model of incompressible turbulence, $t_D \to \infty$ and $\gamma \to 0$, so we recover the Kolmogoroff result, $p_1 = \tfrac{1}{3}$. Second, in the supersonic regime of molecular clouds we expect that there is dissipation by shock waves throughout the cascade, leading to $\gamma > 0$. This implies that the value of the observable exponent $p_1$ is an indicator of dissipation in the supersonic range. According to (47), the assumption that $M \simeq M_c$ forces $\gamma = \frac{1}{2}$. Given this value, we conclude from (49) that the necessary value of $t_D$ is

$$t_D = 0.58 \frac{L}{\sigma_{3D}}. \tag{50}$$

This should be testable using simulations of supersonic flows driven solely by self gravitation, which are not yet available. A number of simulations of nonmagnetic and supersonic turbulence driven by an imposed spectrum of forcing have been carried out over the last decade, a recent example being that by Lemaster and Stone (2009). They find that

$$t_D = 0.78 \frac{\lambda}{\sigma_{3D}}, \tag{51}$$

where $\lambda$ is the wavelength at the maximum of the spectrum of the forcing. Critically important, however, is the fact that the Lemaster – Stone simulations do not include self gravitation, which is an essential part of our model. If and when simulations including self gravitation are carried out, it is reasonable to suppose that the role of $\lambda$ in (51) will be taken over by some parameter proportional to the scale $L$ in our model of the gravitational cascade. If that conjecture proves to be correct, it would support our proposal that $\gamma$ is independent of $L$, so that our solution for $\varepsilon$ in (43) is valid.

*References*

Ballesteros – Paredes, J. 2006, MNRAS, 372, 443
Bertoldi, F., McKee, C.F. 1992, ApJ, 395, 140

Bonnor, W.B. 1956, MNRAS,116, 351

Chieze, J. P. 1987, A&A,171, 225

Cubick, M., Stutzki, J., Ossenkopf, V., Kramer, C., Roellig, M. 2008, A&A, 488, 623

Draine, B., Bertoldi. F. 1996, ApJ, 468, 269

Ebert, R. 1957, ZA, 42, 263

Elmegreen, B. 1989, ApJ, 338, 178

Elmegreen, B., Scalo, J. 2004, AR&AA, 42, 211

Field, G., Blackman, E., Keto, E. 2007, MNRAS,385,181 (FBK)

Field, G., Blackman, E., Keto, E. 2010, in preparation

Gődel, M., Briggs, K., Montmerle, T., Audard, M., Rebull, L., Skinner,S. 2008,Science, 319, 309

Habing, H. 1968, Bull. Astron. Inst. Netherlands,19, 421

Heyer, M.H., & Brunt, C.M. 2004, ApJ,615,L45

Hollenbach, D., Tielens, A. 1999, RevMP, 71,173

Jenkins, E., Tripp, T. 2007, in SINS – Small Ionized and Neutral Structure in the Diffuse ISM, ed. M Haverkorn & W. M. Goss, (San Francisco: Astron. Soc. Pacific ASP Conference Series 365), 51

Keto, E., Myers, P. 1996, ApJ, 304,466

Kolmogorov, A. 1941, Dokl. Akad. Nauk. SSSR., 30,301

Kulesa, C., Hungerford, A., Walker, C., Zhang, X., Lane, A. 2005, ApJ, 625,194

Lemaster, M., Stone, J. 2009, ApJ,691,1092

McKee, C., Ostriker, J. 1977, ApJ, 218,148

Mouschovias, T., Spitzer, L., Jr. 1976, ApJ, 210,326

Pineda, J., and 32 coauthors 2008, A&A, 482,197

Schneider, N., Stutzki, J., Winnewisser, G., Poeglitsch, A., Madden, S. 1998, A&A, 338, 262

Stephens, T., Dalgarno, A. 1973, ApJ,186,165

Stutzki,J., Stacey, G.,Genzel, R., Harris, A., Jaffe, D., Lugten, J. 1988, ApJ, 332,397

Sun, K., Ossenkopf, V., Kramer, C., Mookerjea, B., Roellig, M., Cubick, M., Wannier, P., Andersson, B. – G., Penprase, B., Federman, S. 1999, ApJ, 510, 291

Vazquez – Semadini, E., Gomez, G., Jeppsen, A., Ballesteros – Paredes, J., Gonzalez, R., Klessen, R. 2007, ApJ, 657, 870

Vazquez – Semadini, E., Gomez, G., Jeppsen, A., Ballesteros – Paredes, J., Klessen, R. 2009, ApJ, 707,1023

Whitworth, A., Summers, D. 1985, MNRAS, 214,1

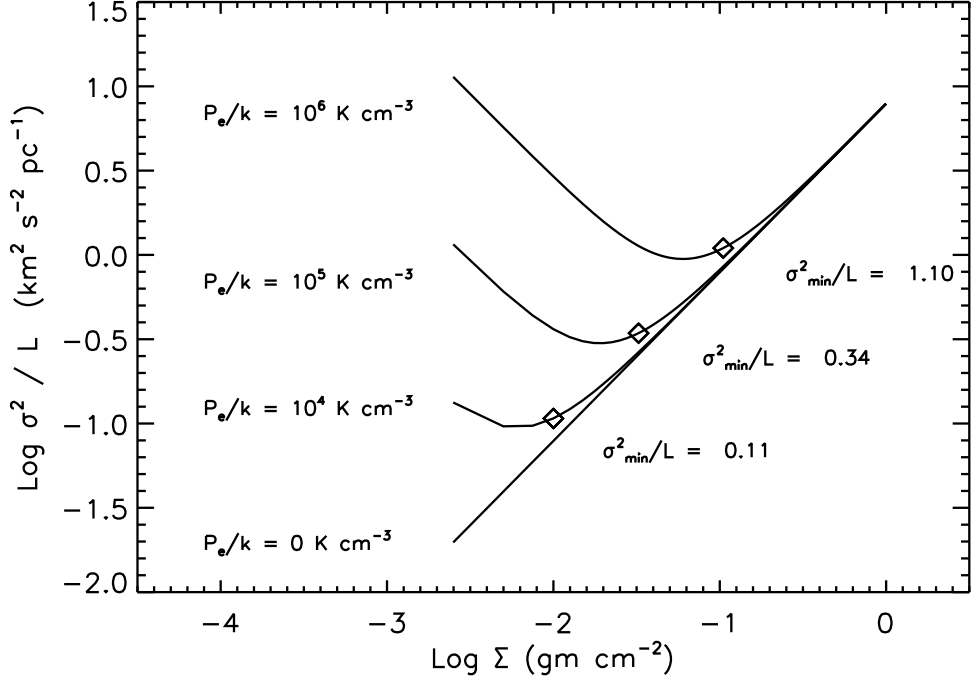

Figure 1. The values of $\sigma^2/L$ derived from Eq. 1 for various column densities $\Sigma$ and external pressures $P_e$. Each solution of constant $P_e$ has a minimum value of $\sigma^2$, marked by the squares, which corresponds to spheres of mass equal to the critical value in Eq. 9, $M_c$. These represent states of marginal stability as discussed in the text. Fragments for which the column density and the quantity $\sigma^2/L$ can be measured are predicted to lie near the squares if they have sufficient mass to achieve criticality, or else on the stable P – branch to the left of the squares if they have insufficient mass and must be pressure confined. Pressure confined stable fragments are thus expected to be smaller. Observable quantities are $N(H_2) = 2.6E24\Sigma$ and $A_V = 250\Sigma$.